| | |
|---|---|
| Title | LDPE surface modifications induced by atmospheric plasma torches with linear and showerhead configurations |
| Authors | S. Abou Rich[1], T. Dufour[1], P. Leroy[1], L. Nittler[2], J-J. Pireaux[2], F. Reniers[1] |
| Affiliations | [1]Université Libre de Bruxelles, faculty of Sciences, Analytical and Interfacial Chemistry, CP 255, Bd Triomphe 2, 1050 Brussels, Belgium<br>sami.abourich@gmail.com, phone : +32 2 650 2994, fax: +32 2 650 2934.<br>[2]University of Namur - PMR, 61 rue de Bruxelles, B-5000, Namur, Belgium |
| Ref. | Plasma Processes & Polymers, 2015, Vol. 12, Issue 8, 771-785 |
| DOI | http://dx.doi.org/10.1002/ppap.201400097 |
| Abstract | Low density polyethylene (LDPE) surfaces have been plasma-modified to improve their nanostructural and wettability properties. These modifications can significantly improve the deposition of subsequent layers such as films with specific barrier properties. For this purpose, we compare the treatments induced by two atmospheric plasma torches with different configurations (showerhead vs linear). The modifications of LDPE films in terms of chemical surface composition and surface morphology are evidenced by X-ray photoelectron spectroscopy, water contact angles measurements and atomic force microscopy. A comparison between the two post-discharge treatments is achieved for several torch-to-substrate distances (gaps), treatment times and oxygen flow rates in terms of etching rate, roughening rate, diffusion of oxygen into the subsurface and hydrophilicity. By correlating these results with the chemical composition of the post-discharges, we identify and compare the species which are responsible for the chemical surface functionalization, the surface roughening and etching. |

# 1. Introduction

Polymers play an essential role in everyday life due to the remarkable diversity of their properties and to successful research aimed at manufacturing products with desired mechanical and chemical properties, tailored for each specific use [1]. Among them, polyethylene is a non-polar polymer without specific chemical functionalities and used in a broad spectrum of applications such as food packaging, medical, automotive, aerospace and electronics fields [2-4]. In that respect, low density polyethylene (LDPE) can be modified to enhance its wettability properties in order to improve the adherence of subsequent deposited films such as barrier layers for food packaging applications [5].

Plasma treatment is known as a very effective approach to improve the hydrophilicity of polymer surfaces without altering their intrinsic bulk properties towards oxygen and water vapor permeation, since only the outermost atomic layers are modified [6-7]. One may prefer using a plasma operating at atmospheric pressure rather than at low pressure as the constraints linked to a costly high-vacuum system are avoided. Furthermore, the implementation of the process in a continuous production line becomes easier. More specifically, atmospheric flowing post-discharges are particularly suitable for the treatment of soft matter such as polymers since maintaining milder conditions than those of a plasma treatment where the same material would be placed between the two electrodes. Most of the electrons and ions are indeed neutralized in the post-discharge which then mostly supplies excited species and long lifetime radicals.

The activation mechanisms of polyethylene surfaces using plasma treatment are fairly well mastered [8-10]. However, the understanding of the diffusion, etching and grafting mechanisms at the plasma/polymer interface is still not well puzzled out, particularly when using atmospheric plasmas with $O_2$ as reactive gas [11-12].

In previous studies [13-15] we have corroborated the necessity of using oxygen to improve the hydrophilicity of LDPE and we have suggested an anisotropic etching mechanism occurring in an atmospheric flowing $Ar-O_2$ or $He-O_2$ post-discharge. For instance, we have shown that an $Ar-O_2$ flowing post-discharge treatment induces a decrease in aWCA as a function of treatment time, until a plateau around 40° is reached after 30 s. This plateau, suggesting a saturation of the oxygenated functions (three polar functionalities, i.e. C-O, C=O, and O-C=O) on the surface has lead us to study the diffusion of oxygen into the bulk of the film (subsurface) [13]. We consider here as "subsurface" the sample's region located immediately below the outer surface and that is not in direct contact with the gaseous species of the plasma phase. Etching rates (between 2.7 nm/s and 7.3 nm/s) specific to the $Ar-O_2$ plasma treatment conditions have been determined from mass losses measurements (between 10 µg/cm$^2$ and 60 µg/cm$^2$) performed as a function of treatment time [14]. A stronger hydrophilicity is obtained for smaller gaps (lower than 15 mm) where – according to optical emission spectroscopy (OES)

results – the more energetic species could break bonds and either eject fragments from the surface or create surface radicals [14]. The ageing study has shown that the aWCA of the modified surfaces increases towards a common threshold of 83° after 30 days of storage, thus indicating a partial recovery of the native wettability state (aWCA = 94°).

In the case of a pure He($O_2$) post-discharge treatment applied to high-density polyethylene samples, same aWCA values as low as 25° have been obtained, whatever the $O_2$ flow rate [15]. The O radicals have been considered as the main reactive species responsible for this wettability since in the pure He post-discharge, we detected mostly O radicals, OH radicals and He metastable atoms while in a He-$O_2$ post-discharge, O radicals, $O_2$ metastable species and $O_2^+$ ions were evidenced.

In this article, we compare the wettability and nanostructural properties of LDPE surfaces after their exposure to two atmospheric post-discharges generated by two different commercial RF plasma torches [16-21]. The first presents a "linear" configuration and uses helium as carrier gas while the second presents a "showerhead" configuration and is supplied with argon. The modifications of LDPE surfaces are investigated by X-ray electron spectroscopy (XPS), drop shape analysis to measure advancing water contact angles (aWCA), atomic force microscopy (AFM) and mass losses measurements. Those results are correlated with the chemical compositions of the post-discharges to identify and compare the gaseous species participating to the surface modifications mechanisms. A comparison between the two plasma treatments is achieved for several torch-to-substrate distances (gaps), treatment times and oxygen flow rates in terms of hydrophilicity, etching rate, roughening rate and diffusion of oxygen into the LDPE subsurface.

## 2. Experimental set-up

### 2.1. Materials & gases

Transparent LDPE films (37% crystalline) provided by PackOplast-Belgium are used, with a thickness of 40 µm, a surface of 20x20 mm² and a density of 0.93g/cm³. The helium, argon and oxygen gases are provided by the company Air Liquide, under the label Alphagaz™ 1 which certifies the following specifications: $H_2O$ < 3ppm vol., $O_2$ < 2 ppm vol. and $C_nH_m$ < 0.5 ppm vol.

### 2.2. Plasma sources

The LDPE films are treated using two plasma sources from SurfX Technologies LLC: the Atomflo™ 400L-Series (referenced here as "linear plasma torch") [22] and the Atomflo™-AH-250D (referenced here as "showerhead plasma torch") [23].

The two plasma devices present common features, namely an RF generator (27.12 MHz), an auto-tuning matching network and a gas delivery system with two mass-flow controllers to regulate the carrier gas (helium or argon) and the reactive gas (here oxygen) supplying the post-discharge. According to the manufacturer's handbooks, the linear torch must be supplied with helium and the showerhead plasma torch with argon as carrier gas, to ensure proper operation. In each configuration, the upper electrode is biased to the RF voltage while the lower is grounded.

In the case of the linear plasma torch (Figure 1a), the gas mixture (helium with/without oxygen) enters through a tube connected to a rectangular housing (55mm*20mm*80 mm). Inside, the gas is homogenized through two perforated sheets, then flows down around the left and right edges of the upper electrode and passes through a slit in the center of the lower electrode. The plasma is lighted and maintained between these electrodes by applying a RF voltage to the upper electrode while the lower electrode is grounded. Due to the rectangular geometry of the slit presenting a length of 20mm and a width of only 0.8 mm (Figure 1c), this plasma torch is commonly described as a "linear plasma torch" with an output area of 16 mm².

The second plasma torch (Figure 1b) consists of two parallel circular electrodes perforated by 126 holes, each one with a critical diameter of 0.6 mm. A 126-holes circular metallic mesh is located parallel and downstream to the two electrodes to homogenize the flowing post-discharge. The gas flow (argon with/without oxygen) is oriented perpendicularly to the two electrodes. As these holes are arranged to form a circular pattern of 20 mm in diameter (Figure 1d), this torch is commonly called "showerhead plasma torch". The output area of this source is evaluated to 35.6 mm² (126*π*R²).

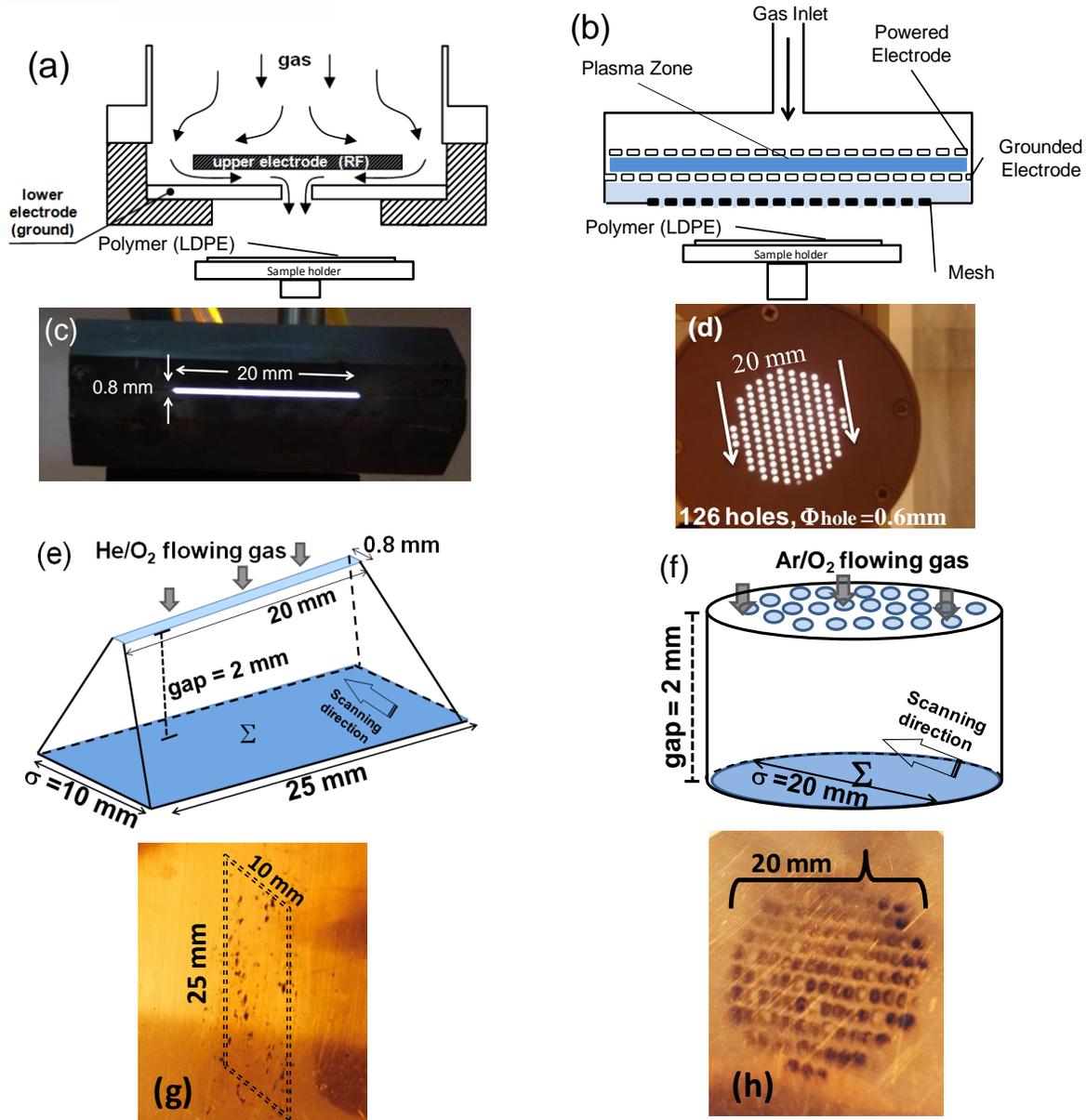

Figure 1. Schematic diagrams of the (a) linear plasma torch (supplied in helium) and (b) showerhead plasma torch (supplied in argon). Pictures of the (c) He post-discharge and (d) Ar post-discharge taken below each torch. Schematic tridimensional profiles of the (e) He post-discharge (linear torch) and (f) Ar post-discharge (showerhead torch). Pictures of a copper plate after exposure to (g) the He post-discharge and (h) the Ar post-discharge.

For all the experiments presented in this article, the plasma torches are used about 10 min after their ignition to reach a stable regime and especially a constant gas temperature. Each of these plasma sources is mounted on a robotic arm enabling its motion back and forth along a single direction and thus the treatment of large surface samples located downstream [24]. In our study, we have set the scanning length ($L_s$) equal to 150mm and the scanning velocity ($v_s$) to 50 mm/s. The distance between the plasma torch and the upper surface of the LDPE sample is called gap and has been varied between 2 and 30 mm. The number of scans ($N_s$) achieved by the robotic arm has been calculated to obtain the same treatment time with the two torches. For instance, a treatment time of 60 s has been achieved with the linear plasma torch for 150 scans while 75 scans were required with the showerhead plasma torch.

Each of these plasma sources generates a flowing post-discharge with specific features. The characteristic width ($\sigma$) of the post-discharge corresponds to its dimension parallel to the scanning direction for a specific value of the gap. Thus, for a gap of 2 mm, $\sigma$ is evaluated to 10 mm for the linear plasma torch and to 20 mm for the showerhead plasma torch, as indicated in Figures 1e and 1f. These values have been evaluated by placing a copper plate 2 mm downstream from the He and Ar flowing post-discharge supplied in oxygen and by measuring the dimensions of the two oxidized patterns as depicted in Figures 1g and 1h. A 10*25mm$^2$ rectangular pattern and a $\pi$*(20mm)$^2$/4 circular pattern have been obtained with the linear and showerhead plasma torches respectively. The area of the post-discharge in contact with the sample ($\Sigma$) depends also on the gap and is represented in Figures 1e and 1f. With a gap fixed at 2mm, $\Sigma$ is equal to 250 mm$^2$ and 314 mm$^2$ for the He and the Ar plasma torches respectively. As the RF power applied to the two torches is 90W, the surface plasma power density estimated on $\Sigma$ is 0.36 W/mm$^2$ for the linear plasma torch and 0.28 W/mm$^2$ for the showerhead plasma torch. As explained in earlier works [24], this difference can be considered as negligible since plasma power variations over this range does not induce significant surface modification in terms of aWCA, mass losses, roughness or chemical surface composition [13]. Those values are reported in Table 1.

|  | Parameters & Variables | Unit | Linear plasma torch | Showerhead plasma torch |
|---|---|---|---|---|
| **Parameters** | Carrier gas nature | - | Helium | Argon |
|  | Output area | mm$^2$ | 16 | 35.61 |
|  | Flow rate | L.min$^{-1}$ | 15 | 33.37 |
|  | Flux | L.min$^{-1}$.mm$^{-2}$ | 0.94 | 0.94 |
|  | Flow velocity | m.s$^{-1}$ | 15 | 15 |
|  | Power | W | 90 | 90 |
|  | $\sigma$ | mm | 10 | 20 |
|  | $\Sigma$ (gap=2 mm) | mm$^2$ | 250 | 314 |
|  | Plasma power density on $\Sigma$ | W.mm$^{-2}$ | 0.36 | 0.28 |
| **Variables** | Treatment time | s | 1-60 | 1-60 |
|  | Number of scans (N$_S$) | - | 1-150 | 1-75 |
|  | O$_2$ flow rate | mL.min$^{-1}$ | 0-40 | 0-80 |
|  | Gap | mm | 2-30 | 2-30 |

*Table 1. Parameters and variables of the linear and showerhead plasma torches used for the treatment of LDPE.*

Another feature of these flowing post-discharges concerns the flow rates which have to vary on similar ranges while respecting the manufacturer limitations, as reported in Table 2. The vector gas flow rates have been fixed at 15 L/min in the case of helium and 33.4 L/min in the case of argon so as to maintain a velocity of 15 m/s at the slit in both configurations for the sake of comparison (see Table 1). The O$_2$ flow rate has been varied from 0 to 40 mL.min$^{-1}$ (helium post-discharge) and from 0 to 80 mL.min$^{-1}$ (argon post-discharge) so as to keep the same flux (0.94 L/min/mm$^2$) in both configurations, as reported in Table 1. The influence of the reactive gas can therefore be estimated from one torch to the other based on the volume fraction of oxygen ($\chi_{O2}$), i.e. the ratio of the O$_2$ flow rate by the carrier gas glow rate. For instance, the value $\chi_{O2}$=0.04/15=2.6.10$^{-3}$ for the He-O$_2$ post-discharge with 40 mL/min of O$_2$ can be compared with the value $\chi_{O2}$=0.08/33.4=2.6.10$^{-3}$ obtained for the Ar-O$_2$ post-discharge with 80 mL/min of O$_2$.

| Parameters | Linear plasma torch | Showerhead plasma torch |
|---|---|---|
| Nature of the carrier gas | Helium | Argon |
| Nature of the reactive gas | Oxygen | Oxygen |
| Carrier gas flow rate range (L/min) | 10-20 | 10-40 |
| Reactive gas flow rate range (mL/min) | 0-800 | 0-80 |
| Power range (W) | 60-160 | 60-110 |

*Table 2. Limitations of the two plasma sources imposed by the manufacturer.*

## 2.3. Diagnostics

The XPS measurements have been performed with a PHI 5600 photoelectron spectrometer, operating at 300 W with a Mg K$_\alpha$ X-ray source (1253.6 eV), under a vacuum of $9.10^{-9}$ Torr, with an angular acceptance of 60° and a number of angular channels of 120. Pass energies of the survey spectra and high resolution spectra have been fixed at 93.90 eV and 23.5 eV respectively. The take-off angle (TOA) of the photoelectrons is 45° with respect to the sample normal axis. The C 1s core level at 285.0 eV has been used to calibrate the binding energy scale. The surface elemental composition has been calculated after removal of a Shirley background by using the following sensitivity coefficients: $S_C = 1$ and $S_O = 2.85$.

To carry out ARXPS (Angle Resolved X-ray Photoelectron Spectroscopy) measurements, the TOA defined between the axis of the photoelectron collection lens and the sample plane has been varied by steps of 10° as follows: 15°, 25°, 35°, 45°, 55°, 65°, 75° [13,25]. In a film of slightly oxidized polyethylene, the photoelectron emitted by an O 1s atom presents a kinetic energy of ≈721 eV corresponding to an attenuation length (λ) of 2.5 nm. In the case of a C 1s atom, this electron kinetic energy is estimated to ≈969 eV, which corresponds to λ = 3.1 nm [26]. For TOA = 90°, the average XPS analysis depth from which ≈63% of the signal originates could be calculated by considering a single attenuation length. But it is also well established that 95% of the information obtained by XPS comes from within three attenuation lengths of the surface (3λ). For this reason, the analysis depth (z) is calculated from the following equation, according to which z = 2.8λ for TOA = 75°, while z = 0.8λ for TOA = 15° [27, 28]. The Table 3 represents the analysis depths (nm) calculated from the equation by considering that the intensity on the top surface (with TOA equal to 15°) is the reference intensity, since the XPS analyser limits are 15° and 75° [29,30]:

$$z(nm) = 3 \cdot \lambda \cdot \sin(TOA)$$

| TOA | | 15° | 25° | 35° | 45° | 55° | 65° | 75° |
|---|---|---|---|---|---|---|---|---|
| z (nm) | O 1s | 2.0 | 3.3 | 4.4 | 5.5 | 6.4 | 7.1 | 7.6 |
|  | C 1s | 2.4 | 3.9 | 5.3 | 6.5 | 7.5 | 8.3 | 8.9 |

*Table 3. Sampling depth analysis (nm).*

The peak fitting of the C 1s components is computed with the Casa XPS software. The bond energies of the C-C, C-O, C=O and O-C=O components have been fixed to 285.0 eV, 286.5 eV, 288.0 eV and 288.9 eV respectively and the FWHMs of each component have been set between 1.5-1.8 eV.

Water contact angle (WCA) measurements have been performed using a drop shape analyzer (Krüss DSA 100) in an air-conditioned room and using milli-Q water as working liquid. Advancing (aWCA) and receding (rWCA) angles have been measured by depositing and withdrawing a droplet of 5 µL on the surface. In this article, each value of dynamic WCA corresponds to the average of 5 drops measurements, randomly deposited onto the sample surface.

The mass losses of the plasma-treated LDPE films have been evaluated by employing a Sartorius BA110S Basic series analytical balance, characterized by a 110 g capacity and 0.01mg precision.

The surface roughness has been evaluated using atomic force microscopy (AFM). The device is a Dimension 3100 AFM using a Nanoscope IIIa controller equipped with a phase imaging extender, from Digital Instruments operating in the Tapping-Mode (TM-AFM). Standard silicon tips (Tap300Al, BudgetSensors) with a 42 N/m nominal spring constant and a 300 kHz nominal resonance frequency have been used. All images (5x5 µm² scanning area) have been recorded in air at room temperature with a scan velocity of 1 Hz. Except a second order polynomial function background slope correction, no further filtering has been performed. From these flattened corrected data, the root-mean-squared roughness ($R_{rms}$) and the maximum topographic height are determined on the flattened 5x5 µm² images.

## 3. Results

The treatments of LDPE surfaces induced either by a He (with/without $O_2$) or an Ar (with/without $O_2$) flowing post-discharge are compared in four dedicated sections dealing with the influence of the gap (g), the treatment time (t), the addition of reactive gas ($O_2$) and the diffusion of functional groups in the subsurface.

### 3.1. Influence of the treatment time

The wettability state of a polyethylene surface exposed to a cold plasma generally depends on the treatment time. Figure 2.a illustrates the variation of the aWCA versus the treatment time for a gap of 2 mm and 90W of plasma power. A decrease from 94° (untreated LDPE sample) to 37° and 31° is evidenced with the showerhead (Ar) and linear (He) plasma torches respectively.

These measurements can be correlated with the O/C ratio where O and C correspond to the relative surface composition of oxygen and carbon determined from the O 1s (532 eV) and C 1s (285 eV) peaks measured by XPS. Figure 2b shows a slight increase in the O/C ratio with the treatment time, thus indicating a gradual enrichment of the surface in oxygen functionalities. The O/C ratios are slightly higher when using the linear plasma torch with helium and as a result the wettability as well. Whatever the plasma torch used, it is noticeable that the O/C ratio tends to reach a plateau for treatment times higher than 30 seconds.

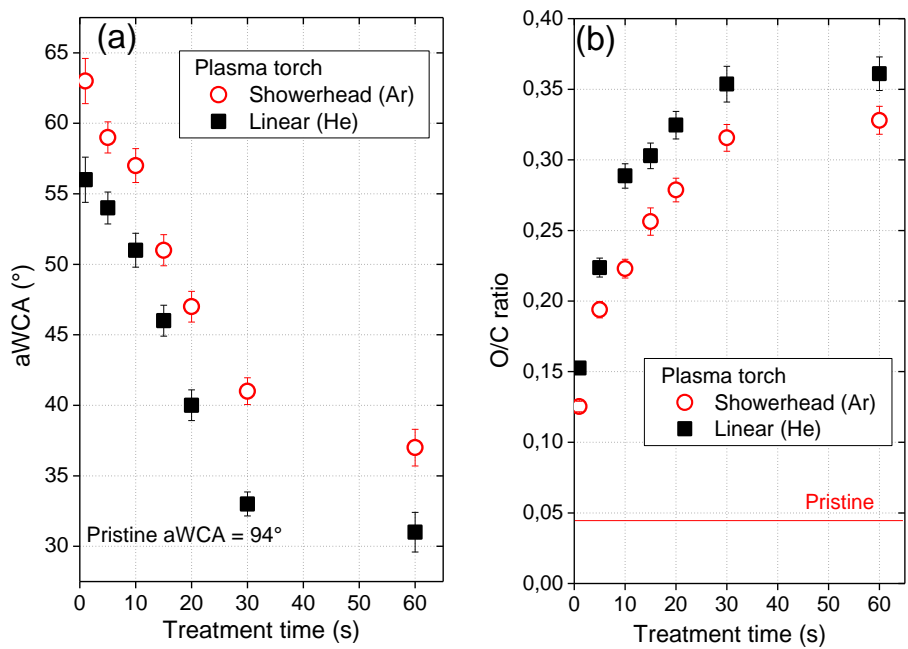

*Figure 2. (a) Variation of the aWCA and (b) O/C ratio of LDPE surfaces exposed to the linear (He) and showerhead (Ar) plasma torches without the addition of oxygen, vs. the treatment time between 1 s and 60 s, $P_{RF}$=90W, g=2mm.*

Figure 3.a shows typical mass losses of LDPE samples after their exposure to a post-discharge as a function of the treatment time. The mass losses seem to vary linearly with the plasma exposure time, since a linear fit is obtained with a correlation coefficient of 0.985. As the density of LDPE ($\rho_{LDPE}$) is 0.93 g/cm$^3$, mass losses (expressed in µg/cm$^2$) can easily be converted into an etching rate (expressed in nm/s) defined as $e_{rate} = \frac{1}{\rho_{LDPE}} \cdot \left(\frac{\text{mass loss}}{t}\right)$. This averaged rate is evaluated to 9.0 nm/s and 13.9 nm/s after a treatment induced by the linear and showerhead plasma torches respectively.

$R_{rms}$ values are reported as a function of the treatment time in Figure 3.b. The $R_{rms}$ values are always much higher than the $R_{rms}$ of the native surface which is only 2 ± 0.3 nm. A linear fit calculated on the data points indicates a correlation coefficient close to 0.977, thus allowing the expression of an averaged roughening rate for these treatment times defined as the ratio of the roughness by its corresponding treatment time. This rate is equal to 0.94 nm/s (linear plasma torch with He) and 1.24 nm/s (showerhead plasma torch with Ar). AFM images of an untreated LDPE surface and several plasma-treated LDPE surfaces for different treatment times are also shown in Figure 3c.

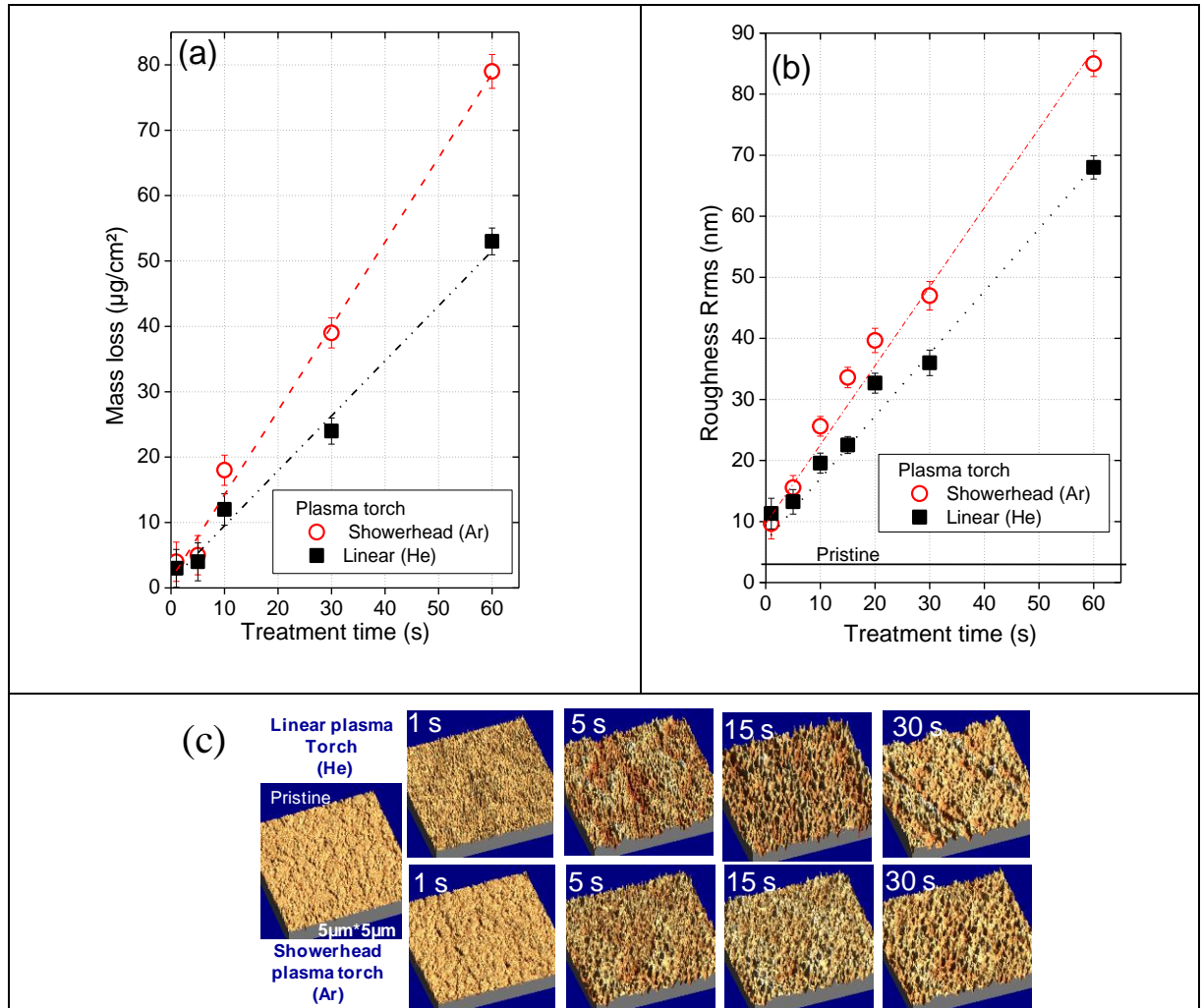

Figure 3. (a) Mass losses, (b) Root mean squared roughness ($R_{rms}$) and (c) Tridimensional AFM images (5*5µm²) vs. treatment time between 1 s and 60 s, using linear (He) and showerhead (Ar) plasma torches (g = 2 mm, $P_{RF}$=90W).

## 3.2. Influence of the gap

The gap is a parameter of interest to determine (i) the critical value beyond which the treatment is no more efficient and (ii) whether impurity sources from the ambient air could contaminate the surface of the treated samples. Figure 4a introduces the values of aWCA as a function of the gap comprised between 2 and 30 mm using the linear and showerhead plasma torches. The two trends are similar since the aWCA increase from 30° to 94° with the linear plasma torch and from 40° to 94° with the showerhead plasma torch. The plateau at 94° obtained for gaps higher than 15 mm corresponds to the native LDPE aWCA and therefore to the fact that the treatment is no more efficient. Below the critical gap of 15 mm, an offset of approximately 8° appears between the two sets of measurements, indicating a treatment more efficient if achieved with the linear plasma torch.

Whatever the configuration of the plasma torch, Figure 4b indicates a decrease in the O/C ratio versus the gap, hence indicating that more oxygen functionalities are grafted on the surface for the shorter gaps. In accordance with Figure 4a, a slight offset between the two curves arises for gaps equal or lower than 15 mm, meaning that the linear plasma torch (pure He) allows a better oxidation of the LDPE surface and therefore a better hydrophilicity state than the showerhead plasma torch (pure Ar). The existence of a plateau for gaps higher than 15 mm is also consistent with the plateau from Figure 4a.

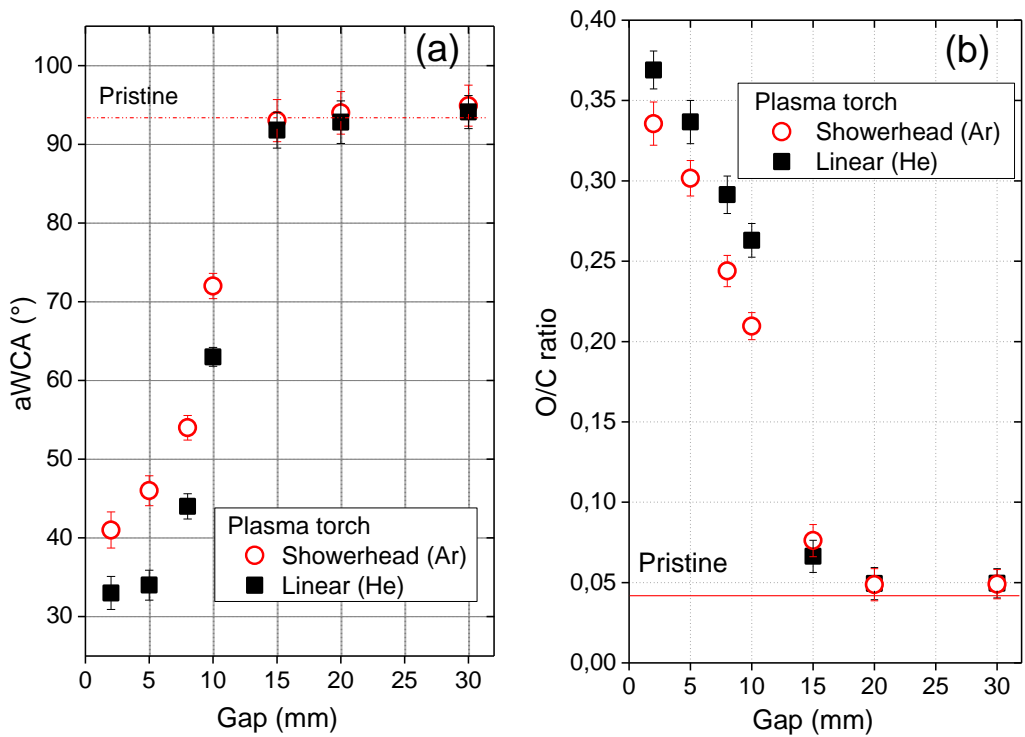

*Figure 4. Variation of (a) aWCA and (b) O/C ratio of LDPE surfaces treated by the linear and showerhead plasma torches vs. the gap for $P_{RF}$=90W, t = 30 s, $\Phi(O_2)$ = 0 mL/min.*

To investigate topographical changes of the plasma-treated surfaces, AFM imaging in tapping mode has been achieved as well as mass losses measurements on the treated samples. The Figure 5a introduces the mass losses of the plasma-treated samples and their corresponding etching rate as a function of the gap. For gaps comprised between 2mm and 30 mm, the etching rate decreases from 7.4 nm/s to approximately 0.7 nm/s in the case of the linear plasma torch, while in the showerhead configuration a decrease from 13.9 nm/s to 0.9 nm/s is depicted. It appears that no material has been removed for gaps higher than 15 mm.

These etching rates can be compared to the roughening rates calculated by expressing the ratio of $R_{rms}$ by the treatment time, as shown in Figure 5b where both the roughness and the roughening rate are plotted as a function of the gap. Consistently with the trends depicted in Figure 5a for a gap comprised between 2 and 30 mm, the roughening rate decreases from 1.2 nm/s to 0.1 nm/s (He post-discharge) and from 1.6 nm/s to 0.1 nm/s (Ar post-discharge). Tridimensional AFM images of an untreated LDPE surface and several plasma-treated LDPE surfaces are shown in Figure 5c for different gaps.

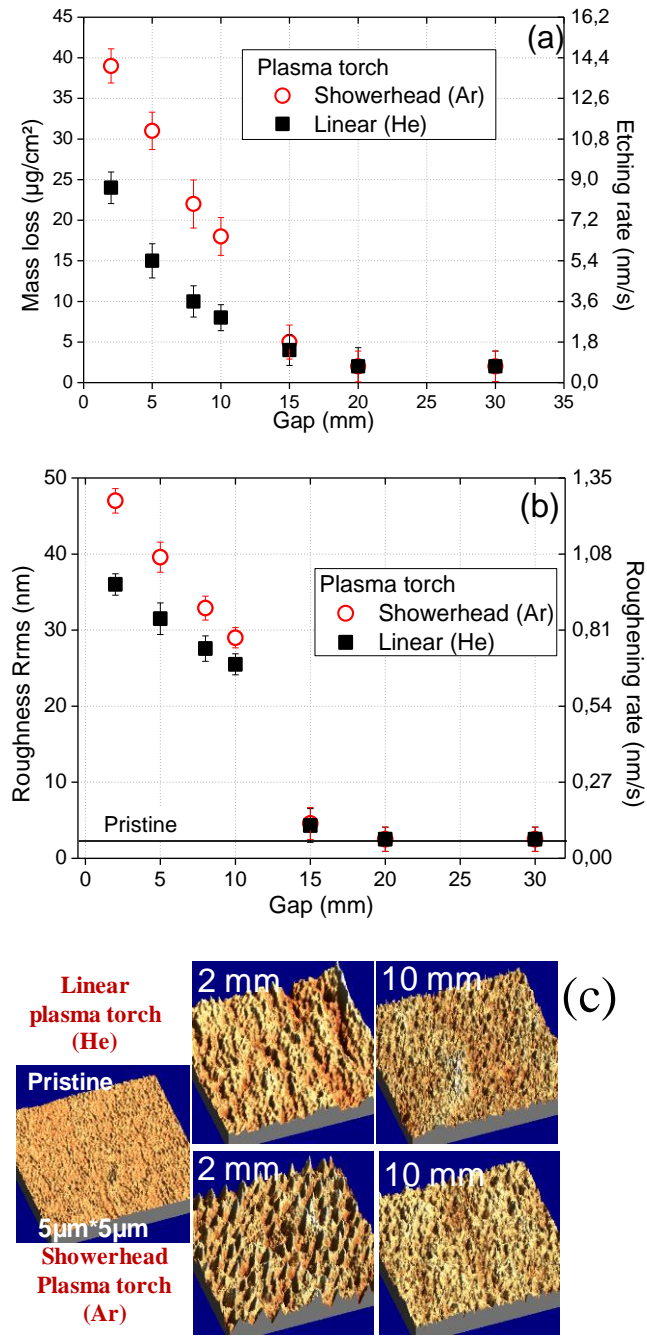

Figure 5. (a) Mass losses and etching rate, (b) Root mean squared roughness ($R_{rms}$) and roughening rate (c) Tridimensional AFM images (5*5µm²) vs. gap between 2nm and 30nm, for treatments achieved with the linear (He) and showerhead (Ar) plasma torches (t = 30s, $P_{RF}$=90W).

## 3.3. Influence of the oxygen volume fraction

Figures 6a and 6b show the variations of the aWCA and the O/C ratios versus the volume fraction of oxygen for He and Ar post-discharge treatments with a gap fixed at 2 mm, an RF power of 90W and a treatment time set to 30 s. Whatever the treatment, the aWCA values decrease until a plateau at about 31° which seems to be reached as soon as $O_2$ is introduced in the post-discharge. However, in absence of this reactive gas, the two wettability states are slightly different, even if much lower than the native wettability state (94°). The O/C ratios from Figure 6b are consistent with the results presented in Figure 6a: the O/C ratios are almost constant and turning around 0.360 when $O_2$ is injected in the plasma torch, while they are slightly lower if no $O_2$ is mixed with the carrier gas: about 0.354 with pure He (in the linear configuration) and 0.336 with pure Ar (in the showerhead configuration).

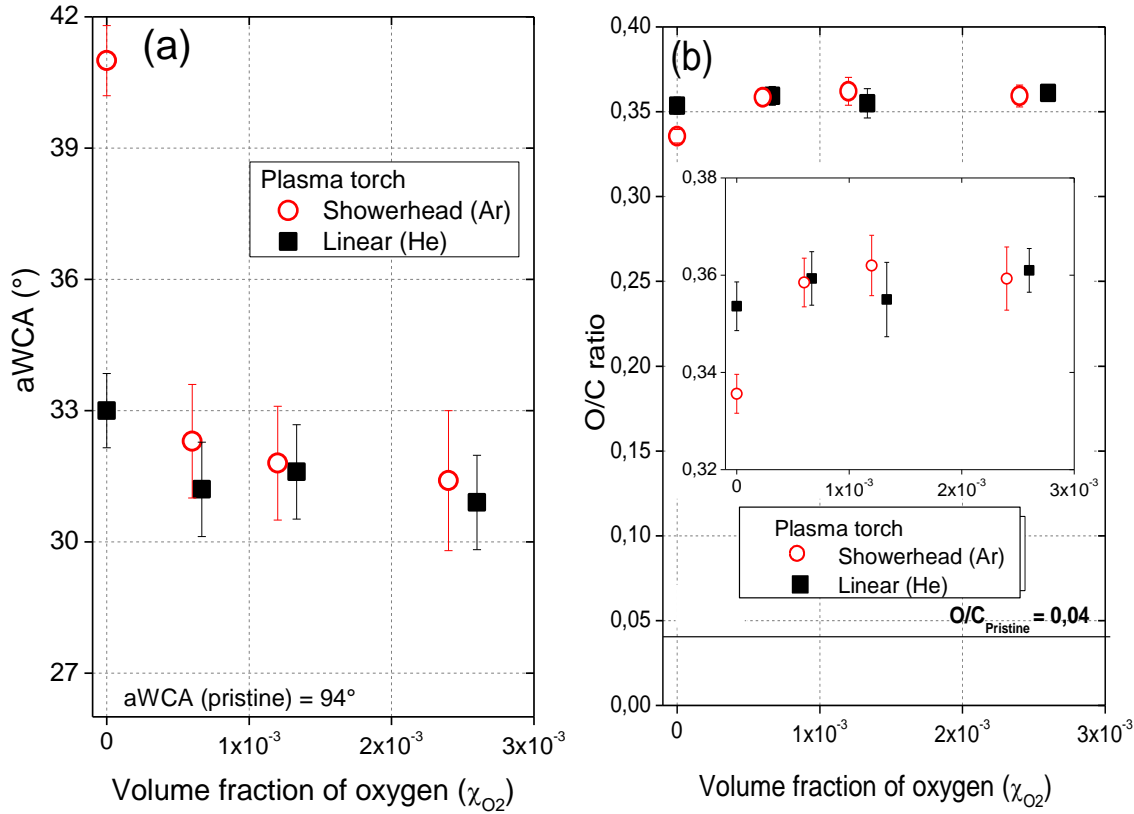

*Figure 6. Variations of the (a) aWCA and (b) O/C ratio vs. volume fraction of oxygen ($P_{RF}$=90W, t =30 s, g=2 mm) for LDPE surfaces treated by the linear and showerhead plasma torches,.*

Figure 7a shows the etching rate (and mass losses) of the sample after its exposure to the post-discharge as a function of the volume fraction of oxygen. The increase of $\chi_{O2}$ from 0 to $2.6 \cdot 10^{-3}$ enhances the etching rate from 9.0 nm/s to 29.5 nm/s (linear plasma torch) and from 13.9 nm/s to 37.3 nm/s (showerhead plasma torch). This raise is correlated with the increase in the roughening rate shown in Figure 7b where on the same $\chi_{O2}$ range, it increases from 0.94 nm/s to 1.86 nm/s (linear plasma torch) and from 1.24 nm/s to 1.89 nm/s (showerhead plasma torch). The Ar post-discharge treatment resulting from the showerhead configuration makes the surface of the films rougher than the alternative treatment.

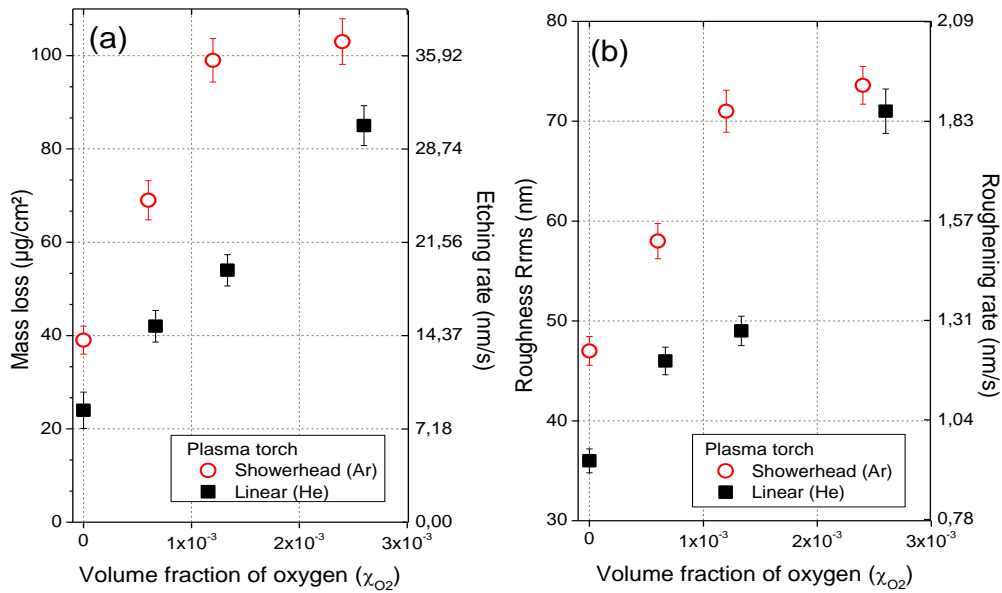

Figure 7. (a) Mass losses and etching rate vs the volume fraction of $O_2$ (b) $R_{rms}$ values and roughening rate vs the volume fraction of $O_2$ for LDPE surfaces treated by the linear and showerhead plasma torches ($P_{RF}$=90W, t = 30 s, g = 2 mm).

## 3.4. LDPE in-depth functionalization

To investigate the in-depth functionalization of plasma-treated LDPE surfaces, ARXPS measurements have been performed on the : C-C, C-O (ether), C=O (carbonyl) and O-C=O (carboxylic) components of the C 1s peak at 285.0 eV, 286.5 eV, 288.0 eV and 289.1 eV respectively. [31]

The profiles of the functional groups are plotted in Figure 8a for the native LDPE surface and in Figures 8b and 8c for the LDPE surfaces exposed to the two plasma treatments with $P_{RF}$=90 W, t=30 s, g=2 mm. The profiles of the two treated surfaces are very different from the profile of the native surface. First, the C-C bonds stand for approximately 92% of the native surface versus 50% and 60-67 % for the samples treated by the linear and showerhead plasma torches respectively. Second, the plasma treatment (Ar or He) induces the production of O-C=O groups which do not exist in the native subsurface. This functional group is lower than 10% on the whole analysis depth after a treatment with the showerhead plasma torch while approaching 20% with the linear plasma torch. The C=O functional groups are more present after a He treatment (>10%) than an Ar treatment (<10% but always higher than the value of the native LDPE which is 2-3% over the whole analysis depth) LDPE surface.

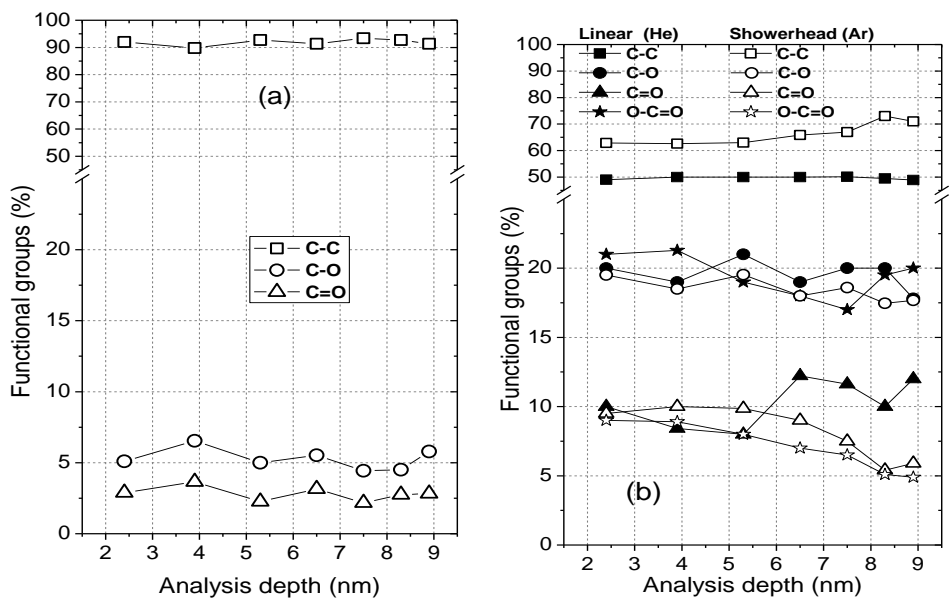

Figure 8. Depth profiles of the C-C, C-O, C=O and O-C=O chemical bonds for (a) native LDPE sample and (b) LDPE samples treated by the linear and showerhead plasma torches ($P_{RF}$=90W, t=30 s, g=2 mm).

## 3.5. Synthetic table

The most relevant results are summarized in Table 4 and discussed in the next section to compare which plasma treatment is the most suitable depending on the desired surface properties. The emissivity values of O and $O_2$ metastable species (in the post-discharge, as measured by OES) have been taken from our previous works [32], [33].

| | Parameters | | Plasma torch configuration | | | | Fig. |
|---|---|---|---|---|---|---|---|
| | | | Linear | | Showerhead | | |
| | | | He | He-$O_2$ | Ar | Ar-$O_2$ | |
| **Plasma** | Oxygen volume fraction $\chi_{O_2}$ | | 0 | $2.6 \cdot 10^{-3}$ | 0 | $2.6 \cdot 10^{-3}$ | |
| | O radicals emissivity (a.u.) | | 1250 | 4400 | 1000 | 3500 | |
| | $O_2$ metastable species emissivity (a.u.) | | 200 | 850 | - | - | |
| **Surface** | Etching rate (nm/s) | Native | 0 | | | | 5a |
| | | Highest value | 7.4 | 29.5 | 13.9 | 37.3 | 7a |
| | Roughening rate (nm/s) | Native | 0 | | | | 5b |
| | | Highest value | 0.94 | 1.86 | 1.24 | 1.88 | 7b |
| | Wettability (°) | Native | 94.0 | | | | 4a |
| | | Lowest value | 33.1 | 30.9 | 40.9 | 31.4 | 6a |
| | O/C ratio | Native | 0.04 | | | | 6b |
| | | Highest value | 0.353 | 0.361 | 0.336 | 0.362 | 7a |
| | In-depth profile of C-O (%) | Native | 5.5 | | | | 8a 8b |
| | | Value at 2.5nm | 20 | - | 20 | - | |
| | | Value at 9nm | 18 | - | 18 | - | |
| | In-depth profile of C=O (%) | Native | 3.2 | | | | |
| | | Value at 2.5nm | 10 | - | 10 | - | |
| | | Value at 9nm | 12 | - | 6 | - | |
| | In-depth profile of O-C=O (%) | Native | 0 | | | | |
| | | Value at 2.5nm | 22 | - | 9 | - | |
| | | Value at 9nm | 20 | - | 5 | - | |

*Table 4. Table synthesizing the main results obtained from a LDPE surface treated by the linear or the showerhead plasma torch with/without oxygen and operating at atmospheric pressure for g=2 mm, $P_{RF}$=90W and t=30s.*

# 4. Discussion

The discussion is elaborated in four sections: first we remind the main species generated by the two torches on the basis of our previous works achieved using optical emission spectroscopy. Then, a correlation between roughening rate and etching rate is presented, followed by a discussion on the relation between wettability and surface activation. Finally, the subsurface functionalization is discussed through the ARXPS measurements.

## 4.1. Reminder on the main species detected by OES in the linear and showerhead plasma torches

The pure He flowing post-discharge generates highly-excited states of He, He metastables, O and OH radicals. Although some VUV are known to be emitted by plasma, the effect of these radiations can be considered as negligible for the treatment of polymers due to the ambient air oxygen absorption [15]. In the pure Ar post-discharge, the main species are highly-excited states of Ar but also OH and O radicals [33-34-35]. Also, simulations performed by Atanasova et al on the showerhead plasma torch show the presence of $Ar_2^+$ ions in the post-discharge few millimeters away from the grounded electrode [36]. Even if these ions could not be directly detected with optical emission spectroscopy, at least a positive current of +200 µA was evidenced in the post-discharge by placing a metallic plate 1 mm downstream of the flow. This current being positive, it represents a flow of positively charged gaseous species that could result from $Ar_2^+$ ions. $O^+$, $N^+$, $O_2^+$, $N_2^+$ or even $Ar^+$ ions might participate to this positive current but as no emission from them could be detected by optical emission spectroscopy, they might be involved in non-radiative processes [33].

Mixing $O_2$ with the He carrier gas consumes the He metastable species to produce $O_2^+$ and $N_2^+$ ions through Penning ionization reactions but also enhances the production of O radicals and $O_2$ metastables [15]. Mixing $O_2$ with the Ar carrier gas leads to the consumption of the Ar metastable species through reactions different from Penning ionizations of $O_2$ since no emission of $O_2^+$ ions could be evidenced. Also, the production of O radicals is strongly enhanced. As the positive current measured in the post-discharge decreases with the $O_2$ flow rate, a consumption of the $Ar_2^+$ ions seems plausible.

## 4.2. Roughening rate & etching rate

The previous results of roughening and etching rates have been synthesized in Table 5 and correlated with the emission of O radicals and $O_2$ metastable species, considering post-discharges with/without oxygen. When no oxygen is mixed with the carrier gas, both the etching and roughening rates become stronger using the showerhead plasma torch. According to the simulations of Atanasova et al [36], the most suitable hypothesis would lie on the presence of the $Ar_2^+$ ions responsible for a physical sputtering. On the contrary, no $He_2^+$ band although observable in the visible range could have been detected. This assumption is sustained by the measurement of positive currents 1 mm away from the plasma torch which are close to 20µA and 200 µA for the linear and showerhead configurations respectively.

Mixing oxygen ($\chi_{O2}$=2.6.10$^{-3}$) with helium in the linear plasma torch enhances the roughening and etching rates. The same scenario occurs for the showerhead plasma torch when $O_2$ is added to Ar. Raisings of these rates seem directly depend on the O radicals production. Indeed and as previously stated, a rise in the $O_2$ flow rate supplying any of the plasma torch increases its dissociation into O radicals. The Table 5 indicates for $\chi_{O2}$=2.6.10$^{-3}$ a stronger dissociation using the linear plasma torch since the emission of O is 4400 a.u. while it is 3500 a.u. for the showerhead configuration. A first explanation may lie on the Penning ionization of $O_2$ which only occurs in presence of He metastables, leading to the production of $O_2^+$ ions. The subsequent dissociation of these ions is a major pathway for the production of O radicals [32]. Also, the highest $O_2$ dissociation using the linear plasma torch could result from a highest plasma power density as reported in Table 1, i.e. 0.35 W.mm$^{-2}$ for the linear plasma torch while 0.28 W.mm$^{-2}$ for the showerhead plasma torch. Besides, $O_2$ metastables solely produced in presence of He and enhanced by increasing the $O_2$ flow rate may participate to the roughening and etching rates as well. Last but not least, the heating of the sample exposed to the post-discharge (and for longer treatment times, e.g. $N_S$=150) may induce a surface melting (liquid phase), giving rise to new micro-patterns appearing as solidified droplets once the post-discharge is switched off.

|  | Linear plasma torch (He) | | | Showerhead plasma torch (Ar) | | |
| --- | --- | --- | --- | --- | --- | --- |
|  | $\chi_{O2}$=0 | $\chi_{O2}$=2.6.10$^{-3}$ | Trend | $\chi_{O2}$=0 | $\chi_{O2}$=2.4.10$^{-3}$ | Trend |
| **O radicals** | 1250 | 4400 | +266% | 1000 | 3500 | +250% |
| **$O_2$ metastable species** | 200 | 850 | +325% | 0 | 0 | - |
| **Roughening rate (nm/s)** | 0.94 | 1.86 | +98% | 1.24 | 1.88 | +52% |
| **Etching rate (nm/s)** | 9.0 | 29.5 | +222% | 13.9 | 37.3 | +168% |

*Table 5. Optical emission of oxygenated species, roughening rate and etching rate of LDPE surfaces treated by the linear and showerhead plasma torches.*

The surface roughness as a function of the mass loss is plotted in Figure 9 in the case of LDPE films treated for different gaps. Two regimes could be distinguished for a fixed treatment time: the first where the roughness increases as a linear function of the mass loss for gaps lower than or equal to 10 mm, and the second regime for gaps higher or equal to 15 mm, where mass losses remain always lower than 5 µg/cm$^2$ and the roughness close to 4 nm. Both plasma treatments seem efficient only for a gap lower than 10 mm, independently of the carrier gas nature and the configuration. This figure clearly indicates that the showerhead plasma torch (Ar) is more suitable to induce structural surface modifications since the roughness and mass losses operate on wider ranges than those. In the framework of our experimental conditions, the surface etching is always accompanied by an increase in its roughness. The absence of a plateau, whether for the roughness or the mass loss, sustains the assumption of two self-dependent phenomena.

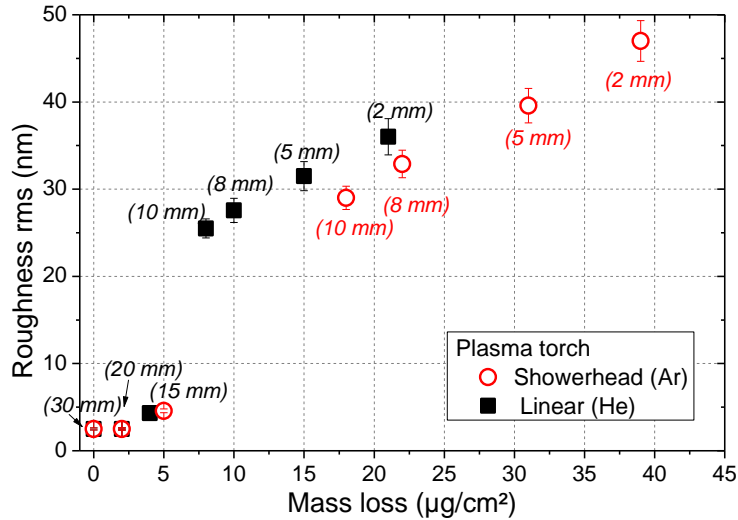

Figure 9. Surface roughness versus the mass loss of LDPE films treated by the linear (pure He) and showerhead (pure Ar) plasma torches, as in the Figure 5 (t=30s, $P_{RF}$=90W). Numbers in parentheses correspond to the values of the gaps.

### 4.3. Wettability and surface activation

The influence of LDPE surface activation on surface wettability can be evidenced by plotting the aWCA as a function of the O/C ratios, as represented in Figure 10. Whatever the plasma torch treatment, the aWCA are significantly lower than the native value at 94.0°.

In presence of oxygen, the nature of the carrier gas plays a minor role on the variations of O/C ratios and as a result on the aWCA values. However, this issue becomes no more negligible in absence of oxygen since aWCA is equal to 33° for a pure helium treatment (linear torch) while 41° for the pure argon treatment (showerhead torch). In the latter case, the highest aWCA value is directly linked to the lowest O/C ratio, namely 0.335. The lowest emissions of O radicals detected in the showerhead configuration are correlated with the lowest oxidation of the treated samples. Besides, the presence of $O_2$ metastables only detected using the linear plasma torch may also be involved in the mechanisms enhancing the oxidation process. On the contrary, the following considerations cannot be taken into account: (i) the carrier gas flux has no influence since it is the same in the two plasma torches, namely 0.94 L/min/mm², as reported in Table 2 (ii) The oxygen impurities from the He and Ar bottles present the same concentrations according to the Air Liquide company. Moreover, they seem negligible (<3 ppm) if compared to the molecular oxygen concentration from the ambient air (approximately 21% mol.).

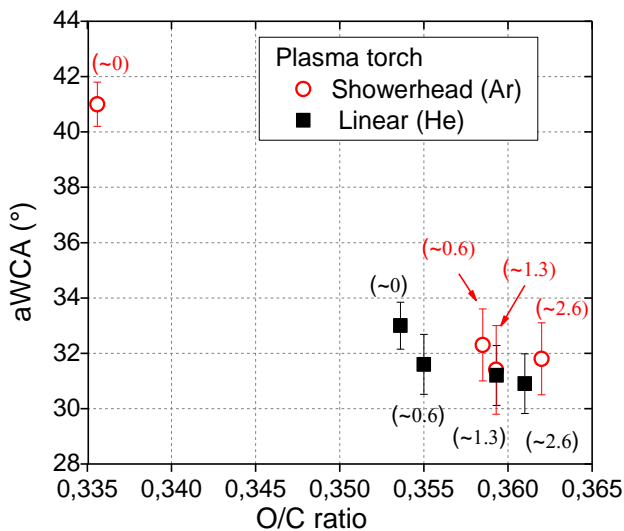

Figure 10. Advancing WCA versus the O/C ratio of LDPE films treated by the linear and showerhead plasma torches, from Figure 6 (t=30s, $P_{RF}$=90W, gap = 2 mm). The numbers in parenthesis correspond to $10^{-3}$ of the $O_2$ volume fraction.

## 4.4. Subsurface functionalization

As a reminder, our pristine LDPE samples are 37% crystalline and 63% amorphous (manufacturer data). The reactive species of the plasma torches are assumed to preferentially etch the LDPE regions which are amorphous rather than crystalline [37]. Then, the diffusion of oxygen atoms into the subsurface may preferentially occur through these amorphous regions (10 times higher than in the crystalline regions [38]), starting from the first microseconds of the plasma treatment and stabilizing 30 s later since the wettability state of LDPE remains unchanged beyond this plasma exposure time. At first sight, the functional groups depth profiles can be considered as surprisingly constant. But if we compare their $R_{RMS}$ parameters (approx.. 40 nm) to the ARXPS analysis depths which is comprised between 0.1 and 9 nm, then the chemical information given by ARXPS corresponds to the upper regions of the rough micro-patterns (or the solidified droplets), i.e. to the upper regions of the LDPE liquid phase that may be formed during the post-discharge exposure. This liquid phase could present a quite homogeneous chemical composition, thus explaining the weak variations in the C-O, C=O and O-C=O profiles.

Regarding the profiles components with more accuracy, the C-O and C=O groups are detected in the same proportions whatever the treatment. However, O-C=O shows a content twice higher with the linear plasma torch (20%) rather than with the showerhead plasma torch (10%). Several assumptions can be drawn: (i) the results of optical emission spectroscopy reported in Table 5 indicate that more O radicals are produced using helium in the linear plasma torch. (ii) The temperature measured on the LDPE surface during the first 60 s was 114°C ± 2°C when using the showerhead plasma torch and 140°C ± 2°C when using the linear plasma torch. As the thermal conductivity of He is 8 times higher than the one of Ar [39], this result seems consistent. Then, the formation of the O-C=O functions could depend on the surface temperature and therefore present higher surface concentrations when using the linear plasma torch with helium. (iii) The formation rate of O-C=O groups could depend on the plasma power density, which is highest in the linear configuration.

For the two treatments, the increasing incorporation of oxygenated functions on the surface could result from several mechanisms whose most significant ones are the following. First, the abstraction of a H atom by an O or OH radical from the surface can lead to the formation of an alkyl radical. As it is unstable in the standard conditions (atmospheric pressure, ambient temperature), it could react with an O atom to form an alkoxy radical (R1). Second, the reaction of a C-radical site with an oxygen radical can lead to the production of a peroxy radical (–C–O–O$^*$), as suggested by the reaction (R2). This product can subsequently react with a polyethylene chain (R) to generate hydroperoxides (C–O–OH) obtained in reaction (R3) [38]. More detailed mechanisms based on the correlation between simulations and ARXPS measurements can be found in a previous study where we exposed LDPE surfaces to a He post-discharge [25].

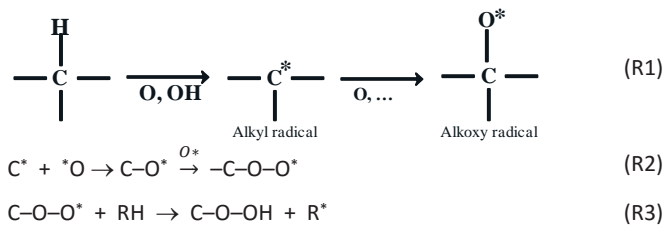 (R1)

$$C^* + {}^*O \rightarrow C-O^* \xrightarrow{O^*} -C-O-O^* \quad (R2)$$

$$C-O-O^* + RH \rightarrow C-O-OH + R^* \quad (R3)$$

## 5. Conclusion

We have compared the treatments of LDPE films induced by two atmospheric flowing post-discharges generated by plasma torches with linear and showerhead configurations and respectively supplied with helium and argon (as carrier gas) . For this purpose, the etching rate, the roughening rate, the chemical surface composition and aWCA measurements of the plasma-treated surfaces have been characterized. For applications focused on the wettability properties it appears that the choice of the torch does not matter if oxygen is added as a reactive gas since in both cases, the aWCA are almost the same and the O/C ratio as well. A fine study shows however that aWCA are slightly lower (about 1-2°) in a He-O$_2$ post-discharge (linear configuration) due to a stronger production of O-C=O functions than in the Ar-O$_2$. Several reasons have advanced in particular a higher plasma power density. On the contrary, the choice of the plasma torch matters if no reactive gas is mixed: using the linear plasma torch enables to reach a more hydrophilic state. For applications more focused on textural modifications of LDPE surfaces, the two plasma torches can induce similar modifications but not to the same extent. In that respect, the surface roughening could result from the melting of the subsurface (liquid phase), giving rise to new micro-patterns appearing as solidified droplets once the post-discharge is switched off. In the framework of our experimental conditions, we have also shown that the surface etching is always correlated with an increasing roughening: no surface roughness plateau has been reached during the ejection of the polymer fragments.

# 6. Acknowledgments

This work was part of Mons-Brussels collaboration supported by the European Commission/Région Wallonne FEDER program (Portefeuille 'Revêtements Fonctionnels'). Research in Mons is also supported by the Science Policy Office of the Belgian Federal Governement (PAI 7/05), the OPTI2MAT Excellence Program of Région Wallonne, and FNRS-FRFC. The authors also thanks the I.A.P/7 (Interuniversitary Attraction Pole) program 'PSI-Physical Chemistry of Plasma Surface Interactions', financially supported by the Belgian Federal Office for Science Policy (BELSPO). The authors gratefully thank Mr. Philippe De Keyzer for his technical support.